\pdfoutput=1
\RequirePackage{fix-cm}
\documentclass[reqno]{amsproc}
\usepackage{amssymb}

\usepackage[citation-order,nobysame]{amsrefs}
\usepackage{xyzbib}

\usepackage{mathtools}

\usepackage{hyperref}

\mathtoolsset{showonlyrefs}

\let\cite=\cites
\newtheorem{theorem}{Theorem}
\newtheorem{proposition}{Proposition}
\newtheorem{lemma}{Lemma}

\DeclareMathOperator{\Asym}{\mathcal{A}}
\DeclareMathOperator*{\Res}{\textrm{Res}}

\newcommand{\rmd}{\mathrm{d}}
\newcommand{\rmi}{\mathrm{i}}

\newcommand{\eps}{\varepsilon}

\let\leq=\leqslant
\let\geq=\geqslant

\newcommand{\psitop}[1]{Z^{\mathrm{top}}_{#1}}
\newcommand{\psibot}[1]{Z^{\mathrm{bot}}_{#1}}

\begin{document}

\title{Integral formulas and antisymmetrization relations
for the six-vertex model}

\author{Luigi Cantini}
\address{LPTM (CNRS UMR 8089), Universit\'e
de Cergy-Pontoise, F-95302 Cergy-Pontoise, France}
\email{luigi.cantini@u-cergy.fr}

\author{Filippo Colomo}
\address{INFN, Sezione di Firenze\\
Via G. Sansone 1, 50019 Sesto Fiorentino (FI), Italy}
\email{colomo@fi.infn.it}

\author{Andrei G. Pronko}
\address{Steklov Mathematical Institute, 
Fontanka 27, 191023 Saint Petersburg, Russia}
\email{agp@pdmi.ras.ru}

\begin{abstract}

We study the relationship between various integral formulas for
nonlocal correlation functions of the six-vertex model with domain
wall boundary conditions. Specifically, we show how the known
representation for the emptiness formation probability can be derived
from that for the so-called row configuration probability.  A crucial
ingredient in the proof is a relation expressing the result of
antisymmetrization of some given function with respect to permutations
in two sets of its variables in terms of the Izergin-Korepin partition
function. This relation generalizes another one 
obtained by Tracy and Widom in the context of the asymmetric simple 
exclusion process.
\end{abstract}

\maketitle

\tableofcontents
\section{Introduction}

In the study of correlation functions of the six-vertex model, and of
the closely related Heisenberg XXZ spin chain, representations in
terms of multiple integrals play an important role
\cite{KBI-93,JM-95,KMST-02,GKS-04}. 
Besides allowing for an exact treatment
\cite{BKNS-02,BKS-03}, and for asymptotic analysis of the
correlation functions \cite{KKMST-09,K-19}, they 
stimulated the development of novel algebraic approaches
\cite{BJMST-06,JMS-09,MS-19}.

For the six-vertex model with domain wall boundary conditions
\cite{K-82,I-87,ICK-92}, which we consider here,
some multiple integral representations are available, for example, for
the emptiness formation probability \cite{CP-07b}. 
They have already proved 
useful in the study of phase separation phenomena in the model, in particular, 
to obtain the arctic curve (frozen boundary of the
limit shape) \cite{CP-09,CP-08}.

To study this model in more detail, for example, in order to
obtain the limit shape (and not just its frozen boundary), more
sensitive correlation functions are necessary. In \cite{CP-12}, the
so-called row configuration probability was introduced, which can be
used as a building block for the calculation of various (both local
and nonlocal) correlation functions. In particular, it can be related
to the emptiness formation probability by certain sum rule.  As
already observed in \cite{CP-12}, to be verified at the level of the
multiple integrals, this sum rule requires a rather non-trivial
identity to hold.

The purpose of the present paper is to expose in detail how integral
formulas for these two correlation functions are connected.  In the
proof, we use essentially a relation expressing the result of
antisymmetrization of some given function with respect to permutations
in two sets of its variables in terms of the Izergin-Korepin partition
function, see Proposition \ref{Cantini}.

It is to be mentioned that similar relations already appeared in the
context of the six-vertex model, see, e.g., \cite{KMST-02}, as well as
in the theory of symmetric polynomials
\cite{KN-99,W-08,BW-16,BWZj-15}.  The relation we use here does not
seem to be a particular case of any of them, even if sharing the
property that its right-hand side is expressible in terms of the
Izergin-Korepin partition function. Instead, it appears to be a
generalization of the antisymmetrization relation given in
\cite{TW-08}, in the context of the asymmetric simple exclusion
process.

\section{Preliminaries}

We consider the six-vertex model in the standard formulation in terms
of arrows on edges (we follow conventions and notations of the
monograph \cite{B-82} and papers \cite{CP-07b,CP-12}). Our aim is to
discuss the model with \emph{domain wall boundary conditions}.  This
means that, given an $N\times N$ lattice (a square lattice with $N$
horizontal and $N$ vertical lines), all arrows on the external edges
are fixed as follows: on the left and right boundaries they are
outgoing (left and right arrows, respectively), while on the top and
bottom boundaries they are incoming (down and up arrows,
respectively).

In the most general setup (compatible with integrability), the weights
of the model depend on the two sets of spectral parameters,
$\lambda_1,\ldots,\lambda_N$ and $\nu_1,\ldots,\nu_N$, where the
parameters are assumed to be all different within each set.  The
parameters of the first set are assigned to the vertical lines, from
right to left, and the parameters of the second set are assigned to
the horizontal lines, from top to bottom. The weights of the vertex
being at the intersection of $k$-th horizontal line and $j$-th
vertical line are $a_{j k}=a(\lambda_j,\nu_k)$, $b_{j
  k}=b(\lambda_j,\nu_k)$, and $c_{j k}=c(\lambda_j,\nu_k)$, where
\begin{equation}\label{abc}
\begin{split}
a(\lambda,\nu)&=\sin(\lambda-\nu+\eta),
\\
b(\lambda,\nu)&=\sin(\lambda-\nu-\eta),
\\
c(\lambda,\nu)&=\sin2\eta. 
\end{split}
\end{equation}
Here, $\eta$ is the so-called crossing parameter, $\Delta=\cos 2\eta$, 
where $\Delta$ is the invariant of the model \cite{B-82}.
The partition function is denoted by 
$Z_N(\lambda_1,\ldots,\lambda_N;\nu_1,\ldots,\nu_N)$
and it is symmetric under permutations within each set of parameters, due to 
integrability of the model via the Yang-Baxter relation.

The partition function was originally studied by Korepin \cite{K-82}
and Izergin \cite{I-87}, who proved that it admits the following
representation in terms of an $N\times N$ determinant:
\begin{multline}\label{ZN}
Z_N(\lambda_1,\ldots,\lambda_N;\nu_1,\ldots,\nu_N)=
\frac{\prod_{j,k=1}^{N}
a(\lambda_j,\nu_k)b(\lambda_j,\nu_k)}{
\prod_{1\leq j<k\leq N}d(\lambda_k,\lambda_j)d(\nu_j,\nu_k)}
\\ \times
\det_{1\leq j,k \leq N}[\varphi(\lambda_j,\nu_k)].
\end{multline}
Here, 
\begin{equation}\label{matT}
\varphi(\lambda,\nu)=\frac{c(\lambda,\nu)}{a(\lambda,\nu) b(\lambda,\nu)},
\end{equation}
and we also denote
\begin{equation}\label{dfunc}
d(\lambda,\lambda')=\sin(\lambda-\lambda').
\end{equation}

In the present paper we mostly discuss the model in the
\emph{homogeneous limit}, that can be obtained by letting
$\lambda_1,\ldots,\lambda_N \to \lambda$, and $\nu_1,\ldots,\nu_N \to
\nu$, where, furthermore, since the weights then are just functions of
the difference $\lambda-\nu$, we put $\nu=0$. In the limit, the
partition function reads
\begin{equation}\label{ZNhom}
Z_N=\frac{(ab)^{N^2}}
{\prod_{j=1}^{N-1}(j!)^2}\, \det_{1\leq j,k\leq N} 
\big[\varphi^{(j+k-2)}\big],\qquad
\varphi^{(n)}\equiv\partial_{\lambda}^{n}\varphi(\lambda,0).
\end{equation}
For quantities of the homogeneous model we use the shorthand notation
$Z_N\equiv Z_N(\lambda,\ldots,\lambda;0,\ldots,0)$, and $a\equiv
a(\lambda,0)$, etc.  In discussion of the correlation functions, we
mostly use, instead of $\lambda$ and $\eta$, the parameters $t$ and
$\Delta$, which in terms of the Boltzmann weights $a$, $b$, and $c$
read:
\begin{equation}\label{tDelta}
t=\frac{b}{a},\qquad 
\Delta=\frac{a^2+b^2-c^2}{2ab}.
\end{equation}

A detailed exposition of proofs of \eqref{ZN} and \eqref{ZNhom} can be
found in \cite{ICK-92}; alternative derivations are also possible,
see, e.g., \cite{BPZ-02,CP-07b}.

An important quantity in the study of the correlation functions of the
model is the one-point correlation function $H_N^{(r)}$, describing
polarization near the boundary. To be more specific, let us consider
the top boundary where all external (vertical) edges, due to the
domain wall boundary conditions, carry down arrows.  Consider now the
next horizontal row of the vertical edges, located between the first
and second horizontal lines. Here, among the $N$ arrows there are
\emph{exactly} $N-1$ down arrows and there is just one up arrow.  The
function $H_N^{(r)}$ gives the probability that this up arrow is
located at the $r$-th vertical edge (recall, that we count vertical
lines from right to left).  Note, that since there is exactly one up
arrow between the first and second horizontal lines, the following sum
rule is valid:
\begin{equation}\label{sumrule}
\sum_{r=1}^{N}H_N^{(r)}=1.
\end{equation}
It is convenient to introduce the corresponding generating function:
\begin{equation}\label{hNz}
h_N(z)=\sum_{r=1}^{N} H_N^{(r)}z^{r-1}.
\end{equation}
Due to \eqref{sumrule}, $h_N(1)=1$. 

Further, we introduce functions $h_{N,s}(z_1,\dots,z_s)$, where the
second subscript, $s=1,\dots,N$, refers to the number of
arguments. These functions are defined as
\begin{equation}\label{hNs}
h_{N,s}(z_1,\dots,z_s) =\frac{1}{
\prod_{1\leq j<k \leq s}^{} (z_k-z_j)}
\det_{1\leq j,k\leq s}\big[
z_k^{s-j}(z_k-1)^{j-1} h_{N-j+1}(z_k)
\big].
\end{equation}
Note that these functions are symmetric polynomials of degree $N-1$ in
each of their variables, and satisfy the reduction relations
\begin{equation}\label{reduction}
  h_{N,s}(z_1,\dots,z_s)\big|_{z_s=1}=h_{N,s-1}(z_1,\dots,z_{s-1}),
  \qquad
  s=1,\ldots,N,  
\end{equation}
with $h_{N,0}\equiv 1$.

In what follows we need the following identity expressing the
partition function of the partially \emph{inhomogeneous} model with
$\nu_j=0$, $j=1,\ldots,N$, in terms of the generating functions of the
boundary correlation functions of the \emph{homogeneous} models on
lattices of the sizes $s\times s$, $s=1,\ldots,N$.
\begin{proposition}
The following representation is valid:
\begin{multline}\label{ZN=hNs}
Z_N(\lambda_1,\dots,\lambda_N;0,\ldots,0)=Z_N(\lambda,\dots,\lambda;0,\ldots,0)
\prod_{j=1}^N\left(\frac{a(\lambda_j,0)}{a(\lambda,0)}\right)^{N-1}
\\ \times
h_{N,N}(\gamma(\lambda_1-\lambda),\dots,\gamma(\lambda_N-\lambda)),
\end{multline}
where the function $\gamma(\xi)$, which also depends on $\lambda$ (and $\eta$)
as a parameter, is given by
\begin{equation}\label{change}
\gamma(\xi)\equiv\gamma(\xi;\lambda)=\frac{a(\lambda,0)}{b(\lambda,0)}
\frac{b(\lambda+\xi,0)}{a(\lambda+\xi,0)}.
\end{equation}
\end{proposition}
The proof can be found in \cite{CP-07b}.

\section{Nonlocal correlation functions}

We discuss here two nonlocal correlation functions. The first one is
the so-called \emph{emptiness formulation probability} (EFP) and it is
denoted as $F_N^{(r,s)}$ \cite{CP-07b}. The name originates from the
spin chain context; in the present model, it gives the probability
that all vertices in an $s \times (N-r)$ rectangular region at the
top-left corner of the $N \times N$ lattice have all the same
configuration of arrows around them. Namely, all these vertices have
left and down arrows, exactly matching the boundary conditions imposed
on the attached boundaries.

The second correlation function is the so-called \emph{row
  configuration probability} (RCP) and it is denoted as
$H_{N,s}^{(r_1,\dots,r_s)}$ \cite{CP-12}. This function gives the
probability of obtaining a given configuration of arrows on all the
$N$ vertical edges located between the $s$-th and $(s+1)$-th
horizontal lines, where there are exactly $s$ up arrows and $N-s$ down
arrows. The integers $r_1,\dots,r_s$ label the positions of these up
arrows, and we set $1\leq r_1<\dots<r_s\leq N$.  The RCP essentially
generalizes the function $H_N^{(r)}$ discussed above to the case of an
arbitrary row.

As indicated in \cite{CP-12}, in dealing with the RCP it is useful to
imagine cutting all the vertical edges between the $s$-th and
$(s+1)$-th horizontal lines of the $N\times N$ lattice, thus
separating the lattice into two parts: the ``top'' one, of size $s
\times N$, and the ``bottom'' one, of size $(N-s)\times N$. The
partition functions of the six-vertex model on these two parts can be
denoted as $\psitop{r_1,\dots,r_s}$ and $\psibot{r_1,\dots,r_s}$,
respectively; clearly,
\begin{equation}\label{defHNs}
H_{N,s}^{(r_1,\dots,r_s)}=\frac{1}{Z_N} \psitop{r_1,\dots,r_s}
\psibot{r_1,\dots,r_s}.
\end{equation}
Note that, although the partition functions $\psitop{r_1,\dots,r_s}$
and $\psibot{r_1,\dots,r_s}$ are in fact very similar objects, the
integers $r_1,\dots,r_s$ play two distinct, complementary, roles in
their definitions.  For $\psitop{r_1,\dots,r_s}$ they indicate the
positions of the arrows at the bottom boundary which are
\emph{reversed} with respect to those on the top boundary (which are
all down arrows).  For $\psibot{r_1,\dots,r_s}$ they indicate the
positions of the arrows at the top boundary which have \emph{the same}
orientation with respect to those on the bottom boundary (which are
all up arrows).

As mentioned in \cite{CP-12}, while the EFP is useful, for example,
for studying phase separation phenomena, and for establishing the
arctic curve \cite{CP-09}, the RCP plays a somewhat more fundamental
role since it can be viewed as a building block to compute other
correlation functions. In particular, the EFP can be expressed in
terms of the RCP, by noting that the EFP can be equivalently defined
as the probability of observing the last $N-r$ arrows between the
$s$-th and $(s+1)$-th horizontal lines to be all pointing down. Since
the positions of the remaining arrows of the row are \emph{not} fixed,
we have
\begin{equation}\label{efp-sumH}
F_N^{(r,s)}=\sum_{1\leq r_1<r_2<\ldots<r_s\leq r}
H_{N,s}^{(r_1,\dots,r_s)}.
\end{equation}

In \cite{CP-12} it was shown that the relation \eqref{efp-sumH}
requires certain cumbersome antisymmetrization relation to be useful
in the context of multiple integral representations.  In the remaining
part of the paper, our aim is to provide all the necessary details
which make it possible to recover the known representation for the EFP
from the those for $\psitop{r_1,\dots,r_s}$ and
$\psibot{r_1,\dots,r_s}$ entering \eqref{defHNs}.

We now recall results for the EFP and RCP in terms of multiple contour
integrals.  In the notation used below, $C_w$ denotes a simple
anticlockwise oriented contour surrounding the point $z=w$ and no
other singularity of the integrand, $\rmd^s z\equiv \rmd z_1 \cdots
\rmd z_s$, and $t\in \mathbb{R}^{+}$ and $\Delta\in \mathbb{R}$ are
the parameters of the homogeneous model, see \eqref{tDelta}.

To elucidate the role of antisymmetrization relations solely on the
example of the EFP, we start with the following result established in
\cite{CP-07b}.
\begin{proposition}\label{EFP}
The EFP admits the representation
\begin{multline}\label{efpMIR1}
F_N^{(r,s)} = (-1)^s
\oint_{C_0}^{} \cdots \oint_{C_0}^{}
\prod_{j=1}^{s}\frac{[(t^2-2\Delta t)z_j+1]^{s-j}}{z_j^r(z_j-1)^{s-j+1}}\,
\\ \times
\prod_{1\leq j<k \leq s}^{} \frac{z_j-z_k}{t^2z_jz_k-2\Delta t z_j+1}\;
h_{N,s}(z_1,\dots,z_s)
\,\frac{\rmd^s z}{(2\pi \rmi)^s},
\end{multline}
where the function $h_{N,s}(z_1,\dots,z_s)$ is defined in \eqref{hNs}.
\end{proposition}
The proof is based essentially on the extensive use of the Yang-Baxter
relation together with methods from the theory of orthogonal
polynomials and properties of the function \eqref{hNz}. We refer to
\cite{CP-07b} for details of the proof.

An apparent drawback of \eqref{efpMIR1} is that the integrand is
\emph{not} symmetric with respect to the permutation of the
integration variables. The central result about the EFP is therefore
the following, which was also established in \cite{CP-07b}.

\begin{theorem}\label{EFPmain}
For the EFP the following representation is valid: 
\begin{multline}\label{efpMIR2}
F_N^{(r,s)}=\frac{(-1)^{s}}{s!}\frac{Z_s}{a^{s(s-1)}c^s}
\oint_{C_0}^{} \cdots \oint_{C_0}^{}
\prod_{j=1}^{s} \frac{1}{z_j^r(z_j-1)u_j^{s-1}}
\prod_{\substack{j,k=1\\ j\ne k}}^{s} \frac{z_k-z_j}{t^2 z_jz_k-2\Delta t z_j +1}
\\ \times
h_{s,s}(u_1,\dots,u_s)
h_{N,s}(z_1,\dots,z_s)
\,\frac{\rmd^s z}{(2\pi \rmi)^s},
\end{multline}
where $Z_s=Z_s(\lambda,\ldots,\lambda;0,\ldots,0)$ is the
Izergin-Korepin partition function of the homogeneous model on
$s\times s$ lattice, and
\begin{equation}\label{uofz}
u_j=-\frac{z_j-1}{(t^2-2\Delta t)z_j+1}, \qquad j=1,\dots,s.
\end{equation}
\end{theorem}

The proof is based on the statement of Proposition \ref{EFP}
\emph{and} certain antisymmetrization relation, which is a special
case of the relation established by Kitanine, Maillet, Slavnov and
Terras in \cite{KMST-02}. We discuss this relation in the next
section.

The representation \eqref{efpMIR2} has played a central role in the
evaluation of the arctic curve of the model \cite{CP-08,CP-09}.

Let us now turn to the integral formulas which determine the RCP, via
the relation \eqref{defHNs}.  We first mention that for
$\psibot{r_1,\dots,r_s}$.
\begin{proposition}\label{Zbot}
For the partition function $\psibot{r_1,\dots,r_s}$
the following representation is valid:
\begin{multline}\label{MIRZ2}
\psibot{r_1,\dots,r_s}=
Z_N\frac{\prod_{j=1}^{s}t^{j-r_j}}{a^{s(N-1)}c^s}
\oint_{C_0}^{} \cdots \oint_{C_0}^{}
\prod_{j=1}^s\frac{1}{z^{r_j}_j}
\prod_{1\leq j <k \leq s}\frac{z_k-z_j}{t^2 z_j z_k -2\Delta t z_j+1}
\\ \times
h_{N,s}(z_1,\dots,z_s)
\, \frac{\rmd^s z}{(2\pi \rmi)^s}.
\end{multline}
\end{proposition}
The proof of this result is again essentially based on the use of the
Yang-Baxter relation together with methods from the theory of
orthogonal polynomials and properties of the function \eqref{hNz}. We
refer to \cite{CP-12} for details.

In turn, for $\psitop{r_1,\dots,r_s}$ we have:
\begin{proposition}\label{Ztop}
For the partition function $\psitop{r_1,\dots,r_s}$ the following
representation is valid:
\begin{multline}\label{oldrep_mir}
\psitop{r_1,\dots,r_s}=
c^s a^{s(N-1)}
\prod_{j=1}^s t^{r_j-j}
\oint_{C_1}^{} \cdots \oint_{C_1}^{} \prod_{j=1}^s \frac{w_j^{r_j-1}}{(w_j-1)^s}\\
\times
\prod_{1\leq j<k\leq s} \left[(w_j-w_k)(t^2 w_j w_k -2\Delta t w_j +1)\right]
\frac{\rmd^s w}{(2\pi \rmi)^s}.
\end{multline}
\end{proposition}
This result follows from the interpretation of
$\psitop{r_1,\dots,r_s}$ as a component of the off-shell Bethe
wave-function when the homogeneous limit is taken \cite{CP-12}.

It is worth emphasizing that the representations \eqref{MIRZ2} and
\eqref{oldrep_mir} are not directly related in any simple way.  For
instance, resorting to the crossing symmetry of the model, one can
indeed easily derive two more representations for
$\psitop{r_1,\dots,r_s}$ and $\psibot{r_1,\dots,r_s}$, but these would
be $(N-s)$-fold (rather than $s$-fold) multiple integrals.

Our aim below is to show how the statement of Theorem \ref{EFPmain}
may be derived from Propositions \ref{Zbot} and \ref{Ztop} via
\eqref{defHNs} and \eqref{efp-sumH}.  It turns out that an essential
ingredient in the derivation is a relation which expresses
antisymmetrization of some expression with respect to permutations of
two sets of variables in terms of the Izergin-Korepin partition
function.

\section{Antisymmetrization relations}\label{antisymm}

Given a function $f(z_1,\dots,z_s)$, introduce the antisymmetrizer
\begin{equation}
\Asym_{z_1,\ldots,z_s}[f(z_1,\dots,z_s)]=
\sum_{\sigma}(-1)^{[\sigma]}
f(z_{\sigma_1},\dots,z_{\sigma_s}), 
\end{equation}
where the sum is taken over permutations $\sigma:1,\dots, s\mapsto
\sigma_1,\dots,\sigma_s$, with $[\sigma]$ denoting the parity of
$\sigma$.

Here, we discuss two antisymmetrization relations which appear to be
of importance for calculation of the EFP. The first one originates
from the following relation established and proved by Kitanine,
Maillet, Slavnov, and Terras in \cite{KMST-02}.
\begin{proposition}[\cite{KMST-02}, Proposition C1] \label{Kitetal}
For the functions $a(\lambda,\nu)$, $b(\lambda,\nu)$ and
$d(\lambda,\lambda')$ given by \eqref{abc} and \eqref{dfunc}, and the
function $e(\lambda,\lambda')=\sin(\lambda-\lambda'+2\eta)$, the
following relation holds
\begin{multline}\label{KMST}
\Asym_{\lambda_1,\ldots,\lambda_s} 
\left[\frac{\prod_{j=1}^{s}\prod_{k=1}^{j-1}a(\lambda_j,\nu_k)
\prod_{k=j+1}^{s}b(\lambda_j,\nu_k)}
{\prod_{1\leq j<k\leq s}^{}e(\lambda_k,\lambda_j)}
\right]
\\
=\frac{\prod_{1\leq j<k\leq s}d(\lambda_k,\lambda_j)}{\prod_{j,k=1}^{s}e(\lambda_k,\lambda_j)}
Z_s(\lambda_1,\ldots,\lambda_s;\nu_1,\ldots,\nu_s),
\end{multline}
where $Z_s(\lambda_1,\ldots,\lambda_s;\nu_1,\ldots,\nu_s)$ is the
Izergin-Korepin partition function \eqref{ZN} for an $s\times s$
lattice.
\end{proposition}

We are interested in the special case of \eqref{KMST}, where
$\nu_j=0$, $j=1,\ldots,s$.  Set
\begin{equation}\label{zjs}
z_j=\gamma(-\lambda_j+\eta), \qquad j=1,\ldots,s,
\end{equation}
where the function $\gamma(\xi)$ is defined in \eqref{change}. We
intend to use \eqref{ZN=hNs}, so it is also useful to introduce the
notation:
\begin{equation}\label{ujs}
u_j=\gamma(\lambda_j-\lambda), \qquad j=1,\ldots,s.
\end{equation}
Note that the variables \eqref{ujs} and \eqref{zjs} are connected by
the relation \eqref{uofz} (recall that $t\equiv
b(\lambda,0)/a(\lambda,0)$ and $\Delta=\cos 2\eta$).  Hence,
\eqref{KMST} for $\nu_j=0$, $j=1,\ldots,s$, can be written, due to
\eqref{ZN=hNs}, as the following antisymmetrization relation:
\begin{multline}\label{identity1}
\Asym_{z_1,\dots,z_s}\left[\prod_{j=1}^{s}\frac{1}{u_j^{s-j}}
\prod_{1\leq j < k \leq s}
(t^2 z_j z_k -2\Delta t z_k+1)
\right]
\\
=(-1)^{\frac{s(s-1)}{2}}\frac{Z_s}{a^{s(s-1)}c^s}
\prod_{1\leq j < k \leq s}(z_k-z_j)
\prod_{j=1}^{s}\frac{1}{u_j^{s-1}}\, h_{s,s}(u_1,\dots,u_s).
\end{multline}
Here (and everywhere below) we assume that $u_j\equiv u(z_j)$, with
the function $u(z_j)$ defined by the right-hand side of \eqref{uofz}.

Clearly, the relation \eqref{identity1} essentially implies the validity
of \eqref{efpMIR2}, given \eqref{efpMIR1}; for more details we refer
to \cite{CP-07b}.

The second antisymmetrization relation appears to be relevant for
obtaining the same result for the EFP from the RCP. It can be
formulated as follows.
\begin{proposition}\label{Cantini}
For $\tau\in \mathbb{C}$, the following relation is valid
\begin{multline}\label{cantini}
\Asym_{x_1,\dots,x_s}\Asym_{y_1,\dots,y_s}\left[
\prod_{j=1}^s  \frac{(x_jy_j)^{s-j}}{1-\prod_{l=1}^j x_ly_l}
\prod_{1\leq j<k\leq s}(x_jx_k+\tau x_k+1)(y_jy_k+\tau y_k+1)
\right]
\\ =\prod_{j,k=1}^{s}(x_j+y_k+\tau x_jy_k)
\det_{1\leq j,k \leq s} \left[\psi(x_j,y_k)\right],
\end{multline}
where 
\begin{equation}
\psi(x,y)=\frac{1}{(1-x y )(x+y+\tau x y)}.
\end{equation}
\end{proposition}
We prove it in Appendix \ref{appa}.

As shown below, the right-hand side of the relation \eqref{cantini} can be
expressed in terms of the Izergin-Korepin partition function. In fact, it
essentially coincides with the right-hand side of some Cauchy-like identity
for the Hall-Littlewood polynomials considered in \cite{BW-16}, see  equation
(25) therein.  As for the left-hand side of the relation (25) in \cite{BW-16},
it can also be rewritten as the result of double antisymmetrization of some
function. However, even in the case $s=2$, the function  under the
antisymmetrization symbol appears to be  different from that standing in
\eqref{cantini}.

By its left-hand side, the relation \eqref{cantini} reminds another
antisymmetrization relation, first established and proved by Tracy and Widom
in \cite{TW-08}, see (1.6) therein, in the context of the asymmetric simple
exclusion process. Indeed, it appears that the relation \eqref{cantini}
reduces to that of \cite{TW-08} in the limit where $y_j\to t^{-1}$,
$j=1,\dots,s$, upon the identification of the parameters $t=\sqrt{q/p}$ and
$\tau=-1/\sqrt{pq}$ in terms of the asymmetric simple  exclusion process rates
$p$ and $q$, $p+q=1$. We give details in Appendix \ref{appb}.

To discuss the relation of \eqref{cantini} with Izergin-Korepin
partition function, it is convenient to introduce the notation
\begin{equation}\label{defW}
W_s(x_1,\dots,x_s;y_1,\dots,y_s)
=\frac{\prod_{j,k=1}^{s}
(x_j+y_k+\tau x_jy_k)}{\prod_{1\leq j<k \leq s}(x_k-x_j)(y_k-y_j)}
\det_{1\leq j,k \leq s}  [\psi(x_j,y_k)].
\end{equation}
Let us identify $\tau=-2\Delta$, and set
\begin{equation}\label{xyCantini}
x_j=\frac{a(\lambda_j,\zeta+\eta)}{b(\lambda_j,\zeta+\eta)}, \qquad
y_j=\frac{a(\zeta,\nu_j)}{b(\zeta,\nu_j)}, \qquad j=1,\dots,s,
\end{equation}
where $\zeta$ is an arbitrary parameter to be fixed later. Then we have
\begin{equation}\nonumber
\psi(x_j,y_k)= \frac{1}{c^3}
[b(\lambda_j,\zeta+\eta)b(\zeta,\nu_k)]^2\varphi(\lambda_j,\nu_k)
\end{equation}
and therefore
\begin{equation}\label{cantiniIK}
\det_{1\leq j,k \leq s}\left[\psi(x_j,y_k)\right]
=\frac{1}{c^{3s}}
\prod_{j=1}^s [b(\lambda_j,\zeta+\eta)b(\zeta,\nu_j)]^2
\det_{1\leq j,k \leq s}\left[\varphi(\lambda_j,\nu_k)\right].
\end{equation}
Note that, in the right-hand side, the parameter $\zeta$ enters only the
prefactor and not the function $\varphi(\lambda_j,\nu_k)$.  Plugging
this into \eqref{defW} yields
\begin{multline}\label{WsIK}
W_s(x_1,\dots,x_s;y_1,\dots,y_s)=
(-1)^s
\frac{\prod_{j=1}^s b(\lambda_j,\zeta+\eta) b(\zeta,\nu_j)}{c^{2s}
\prod_{j,k=1}^{s} b(\lambda_j,\nu_k)} 
\\ \times
Z_s(\lambda_1,\ldots,\lambda_s;\nu_1,\ldots,\nu_s).
\end{multline}

Let us now consider \eqref{WsIK} in the partial homogeneous limit,
where $\nu_j=0$, $j=1,\ldots,s$. To make a contact with our previous
discussion let us, furthermore, identify $\zeta=\lambda$, so that, when
comparing \eqref{xyCantini} with \eqref{zjs}, we have in the limit
\begin{equation}
x_j=t z_j,\quad y_j=t^{-1},\qquad j=1,\ldots,s.
\end{equation}  
Using \eqref{ZN=hNs}, we thus obtain
\begin{equation}\label{Whom}
W_s(t z_1,\dots,t z_s;t^{-1},\dots,t^{-1})
=\frac{(-1)^s Z_s}{c^s b^{s(s-1)}}
\prod_{j=1}^s \frac{1}{(z_j-1)u_j^{s-1}}\,
h_{s,s}(u_1,\dots,u_s).
\end{equation}
Here, as usual, $u_j$'s and $z_j$'s are related by \eqref{uofz}.

\section{From RCP to EFP}

Let us now consider the derivation of \eqref{efpMIR2} from
\eqref{MIRZ2} and \eqref{oldrep_mir}, basing on \eqref{defHNs} and
\eqref{efp-sumH}.  For convenience, we change the integration
variables $z_j\mapsto x_j/t$ in \eqref{MIRZ2}, that yields
\begin{multline}\label{zbotcov}
\psibot{r_1,\dots,r_s}=
\frac{Z_N}{a^{s(N-1)}c^s}
\oint_{C_0}^{} \cdots \oint_{C_0}^{}
\prod_{j=1}^s\frac{1}{x^{r_j}_j}
\prod_{1\leq j <k \leq s}\frac{x_k-x_j}{x_j x_k -2\Delta  x_j+1}
\\ \times
h_{N,s}\left(\frac{x_1}{t},\dots,\frac{x_s}{t}\right)
\frac{\rmd^s x}{(2\pi \rmi)^s},
\end{multline}
and also we change $w_j\mapsto 1/(ty_j)$ in \eqref{oldrep_mir}, 
that yields
\begin{multline}\label{ztopoldcov}
\psitop{r_1,\dots,r_s}
=c^s a^{s(N-1)}
\oint_{C_{1/t}}^{} \cdots \oint_{C_{1/t}}^{} \prod_{j=1}^s
\frac{1}{(ty_j-1)^s y_j^{r_j+s-1}}
\\ \times
\prod_{1\leq j<k\leq s} \left[(y_k-y_j) (y_j y_k -2\Delta y_k+1)\right]
\frac{\rmd^s y}{(2\pi\rmi)^s}.
\end{multline}
Now, putting all together in \eqref{defHNs}, and using
\eqref{efp-sumH}, we have
\begin{multline}\label{efpdoubleMIR}
F_N^{(r,s)}=
\oint_{C_{1/t}}^{} \cdots \oint_{C_{1/t}}^{}
\frac{\rmd^s y}{(2\pi \rmi)^s}
\oint_{C_0}^{} \cdots \oint_{C_0}^{}
\prod_{j=1}^s \frac{1}{y_j^{s-1}(ty_j-1)^s}
\\\times
\prod_{1\leq j<k\leq s}
\frac{(y_k-y_j)(y_j y_k -2\Delta y_k +1)(x_k-x_j)}{x_j x_k -2 \Delta x_j +1}
h_{N,s}\left(\frac{x_1}{t},\dots,\frac{x_s}{t}\right)
\\\times
\sum_{1\leq r_1 < r_2 < \dots < r_s \leq r}
\prod_{j=1}^s \frac{1}{(x_jy_j)^{r_j}}
\frac{\rmd^s x}{(2\pi \rmi)^s}.
\end{multline}
To prove that this representation indeed simplifies to
\eqref{efpMIR2}, one has to: 1) make the multiple sum, 2) perform
symmetrization of the integrand, and 3) evaluate integrations in one
set of variables.

The first task can be accomplished by the following. 
\begin{lemma}\label{ssum}
Let variables $z_1,\ldots,z_s$ all 
take values inside the unit circle in $\mathbb{C}$, that is 
$|z_j|<1$, $j=1,\ldots,s$, then 
\begin{equation}\label{summation}
\sum_{-\infty< r_1 < r_2 < \dots < r_s \leq r}
\prod_{j=1}^s \frac{1}{z_j^{r_j}}
=\prod_{j=1}^s \frac{1}{z_j^{r-s+j}\big(1-\prod_{l=1}^j z_l\big)}.
\end{equation}
\end{lemma}
\begin{proof}
By successively performing the summations in the left-hand side of
\eqref{summation} with respect to $r_1,\ldots,r_s$, in that order, one
arrives at the expression in the right-hand side.
\end{proof}

To apply Lemma \ref{ssum} we first remark that the integral
representation \eqref{zbotcov} vanishes whenever one of the $r_j$'s is
negative.  In \eqref{efpdoubleMIR} we may thus replace the sum over
the values $1\leq r_1<r_2\cdots<r_s\leq r$ by the sum over the values
$-\infty<r_1<r_2\cdots<r _s\leq r$.  Then we can apply
\eqref{summation} by setting $z_j=x_j y_j$, $j=1,\ldots,s$.  As a
result, the expression \eqref{efpdoubleMIR} simplifies to:
\begin{multline}\label{efpdoubleMIR2}
F_N^{(r,s)}=
\oint_{C_{1/t}}^{} \cdots \oint_{C_{1/t}}^{}
\frac{\rmd^s y}{(2\pi \rmi)^s}
\oint_{C_0}^{} \cdots \oint_{C_0}^{}
\prod_{j=1}^s \frac{1}{(ty_j-1)^s y_j^{r+j-1}x_j^{r-s+j}}
\\ \times
\prod_{j=1}^{s}\frac{1}{(1-\prod_{l=1}^{j} x_l y_l)}
\prod_{1\leq j<k\leq s}{}
\frac{(y_k-y_j)(y_j y_k -2\Delta  y_k +1)(x_k-x_j)}{x_j x_k -2 \Delta x_j +1}
\\ \times 
h_{N,s}\left(\frac{x_1}{t},\dots,\frac{x_s}{t}\right)
\frac{\rmd^s x}{(2\pi \rmi)^s}.
\end{multline}

The second step relies on Proposition~\ref{Cantini}. Indeed, the
symmetrization of the integrand of \eqref{efpdoubleMIR2} with respect
to permutations in the two sets of the integration variables
essentially reduce to the left-hand side of \eqref{Cantini} with
$\tau=-2\Delta$.  Using the notation \eqref{defW} for the right-hand
side of \eqref{cantini}, we have
\begin{multline}\label{efpdoubleMIR3}
F_N^{(r,s)}=\frac{1}{(s!)^2}
\oint_{C_{1/t}}^{} \cdots \oint_{C_{1/t}}^{}
\frac{\rmd^s y}{(2\pi \rmi)^s}
\oint_{C_0}^{} \cdots \oint_{C_0}^{}
\prod_{j=1}^s \frac{1}{x_j^r(ty_j-1)^s y_j^{r+s-1}}
\\\times 
\prod_{1\leq j<k\leq s}
\frac{(x_k-x_j)^2(y_k-y_j)^2}{(x_j x_k -2 \Delta  x_j +1)
(x_j x_k -2 \Delta  x_k +1)}
\\ \times
W_s(x_1,\dots,x_s;y_1,\dots,y_s)
h_{N,s}\left(\frac{x_1}{t},\dots,\frac{x_s}{t}\right)
\frac{\rmd^s x}{(2\pi \rmi)^s}.
\end{multline}

At the last step we perform integrals over the variables
$y_1,\dots,y_s$.  For this purpose we use the following.
\begin{lemma}\label{sint}
For an arbitrary symmetric function $\Phi(y_1,\dots,y_s)$,
regular in each of its variables at the point $y=w$, one has
\begin{multline}
\oint_{C_w}\dots\oint_{C_w}
\prod_{j=1}^s\frac{1}{(y_j-w)^s}\prod_{1\leq j<k\leq s}(y_k-y_j)^2
\Phi(y_1,\dots,y_s)\frac{\rmd^s y}
{(2\pi\rmi)^s}
\\
=(-1)^{\frac{s(s-1)}{2}}  s! \Phi(w,\dots,w).
\end{multline}
\end{lemma}
\begin{proof}
In evaluating the residues recursively in each variable, it is easily
seen that only those terms survive where differentiations all act on
the squared Vandermonde product, and the result follows.
\end{proof}

Applying Lemma~\ref{sint} with $w=t^{-1}$, we obtain
\begin{multline}\label{efpMIR2recovered}
F_N^{(r,s)}=\frac{t^{s(r-1)}}{s!}
\oint_{C_0}^{} \cdots \oint_{C_0}^{}
\prod_{j=1}^s \frac{1}{x_j^r}
\prod_{\substack{j,k=1\\ j\ne k}}
\frac{x_j-x_k}{x_j x_k -2 \Delta x_j +1}
\\ \times
W_s(x_1,\dots,x_s;t^{-1},\dots,t^{-1})
h_{N,s}\left(\frac{x_1}{t},\dots,\frac{x_s}{t}\right)
\frac{\rmd^s x}{(2\pi \rmi)^s}.
\end{multline}
Finally, making back the change of the integration variables
$x_j\mapsto t z_j$, $j=1,\dots,s$, and using \eqref{Whom}, we
immediately arrive to \eqref{efpMIR2}.

In conclusion, in the present paper we have studied the relationship between
various integral formulas for nonlocal correlation functions of the six-vertex
model with domain wall boundary conditions. Specifically, we have shown how
the known result for the EFP can be derived from that for the RCP. A crucial
role in the proof has been played by the relation \eqref{cantini}. This
relation may appear useful in the study of other correlation functions of the
model, and it is definitely not purely specific of the domain wall boundary
conditions.  On the other hand, calculation of other correlation functions
than EFP may require the use of different but similar relations, and
establishing them can be the subject of further study.

\section*{Acknowledgments}

We thank the Galileo Galilei Institute for Theoretical Physics (GGI,
Florence), research program on ``Breakdown of ergodicity in isolated quantum
systems: from glassiness to localization'' for hospitality and support at the
final stage of this work. The second author (FC) acknowledges partial support
from MIUR, PRIN grant 2017E44HRF on ``Low-dimensional quantum systems:
theory, experiments and simulations''. The third author (AGP) acknowledges
partial support from the Russian Science Foundation, grant 18-11-00297, and
from INFN, Sezione di Firenze.

\appendix
\numberwithin{equation}{section}
\setcounter{equation}{0}
\section{Proof of  Proposition \ref{Cantini}}\label{appa}

Here we prove Proposition \ref{Cantini}. We apply induction in $s$,
using the symmetries of \eqref{cantini} in the involved variables and
comparing singularities of both sides. Our proof goes in many respects
along the lines of that of Proposition \ref{Kitetal}, given in
\cite{KMST-02} (see Appendix C therein).

Denote by $L_s=L_s(x_1,\ldots,x_s;y_1,\ldots,y_s)$ and
$R_s=R_s(x_1,\ldots,x_s;y_1,\ldots,y_s)$ the left- and right-hand
sides of \eqref{cantini}, respectively. For $s=1$, we immediately
verify that $L_1 =R_1$. Our aim is, assuming $L_{s-1} =R_{s-1}$, to
show that $L_{s}=R_{s}$. 

First we show that $L_s$ can be written in the form
\begin{equation}\label{L-Poly}
L_s=\frac{\prod_{1\leq j<k \leq s}(x_k-x_j)(y_k-y_j)
}{(1-\prod_{i=1}^{s} x_iy_i) \prod_{j,k=1}^{s}(1-x_jy_k)}
P_s(x_1,\ldots,x_s;y_1,\ldots,y_s),
\end{equation}
where $P_s(x_1,\ldots,x_s;y_1,\ldots,y_s)$ is a polynomial of degree
$s$ in each variable, separately symmetric under permutations of the
variables within each set. For this, we split the sums over
permutations in the left-hand side of \eqref{cantini} as follows
\begin{equation}\label{ind-step}
\begin{split}
\left(1-\prod_{i=1}^{s} x_iy_i\right)L_s 
&= \sum_{j,k=1}^{s} \sum_{\substack{\sigma,\rho \\ \sigma_s=j, \rho_s=k
}} (-1)^{[\sigma]+[\rho]}
\prod_{l=1}^{s-1} \frac{(x_{\sigma_l} y_{\rho_l})^{s-l}}
{1-\prod_{i=1}^l x_{\sigma_i}y_{\rho_i}}
\\ &\quad\times
\prod_{1\leq m <n\leq s}
(1+x_{\sigma_m} x_{\sigma_n} + \tau x_{\sigma_n})
(1+y_{\rho_m} y_{\rho_n} + \tau y_{\rho_n})
\\ &
=\sum_{j,k=1}^{s} (-1)^{j+k}
L_{s-1}(\setminus x_j;\setminus y_k)
\\ &\quad\times
\prod_{\substack{m=1 \\ m\neq j}}^{s} x_m(1+x_m x_j+\tau x_j) 
\prod_{\substack{n=1\\n\neq k}}^{s} y_n(1+y_n y_k+\tau y_k).
\end{split}
\end{equation}
Here, $L_{s-1}(\setminus x_j;\setminus y_k)$
denotes the quantity $L_{s-1}$ 
which is constructed from the sets $x_1,\ldots, x_s$ and 
$y_1,\ldots,y_s$ with $x_j$ and $y_k$ removed, respectively.
Noticing that can be written in the form
\begin{equation}\label{R-Poly}
R_{s}
=\frac{\prod_{1\leq j<k \leq s}(x_k-x_j)(y_k-y_j)
}{\prod_{j,k=1}^{s}(1-x_jy_k)}
Q_{s-1}(x_1,\ldots,x_s;y_1,\ldots,y_s),
\end{equation}
where $Q_{s-1}(x_1,\ldots,x_s;y_1,\ldots,y_s)$ is a polynomial of
degree $s-1$ in each variable, separately symmetric under permutations
of the variables within each set, and applying the inductive step
$L_{s-1}(\setminus x_j;\setminus y_k)= R_{s-1}(\setminus x_j;\setminus
y_k)$, we conclude that the expression in \eqref{ind-step} has poles
only at the points $x_j=y_k^{-1}$ and therefore $L_{s}$ can be written
in the form \eqref{L-Poly}.

Next we must show that the expressions \eqref{L-Poly} for $L_{s}$ and
\eqref{R-Poly} for $R_{s}$ are equal; to this purpose, it suffices to
show that they coincide at $s+1$ distinct values of $x_1$.  We show
below that both expressions have equal residues at $x_1=y_j^{-1}$,
$j=1,\dots, s$, and that they coincide for $x_1=0$.

Let us evaluate the residue at $x_1=y_1^{-1}$. For $R_{s}$, we find
\begin{equation}
\Res_{x_1=y_1^{-1}} R_s =-{y_1}^{-1}
\prod_{i=2}^{s} 
(y_1^{-1}+y_i+\tau y_1^{-1}y_i)(x_i+y_1+\tau x_i y_1)
R_{s-1}(\setminus x_1;\setminus y_1). 
\end{equation}
In the case of $L_{s}$, it follows from the antisymmetrization that
the pole at $x_1=y_1^{-1}$ is present only in the terms of the sum in
which $x_1$ and $y_1$ are not permuted. Restricting to these terms and
setting $x_1=y_1^{-1}$, we find
\begin{equation}
\Res_{x_1=y_1^{-1}} L_s=-y_1^{-1}
\prod_{i=2}^s 
(y_1^{-1}+y_i+\tau y_1^{-1}y_i)(x_i+y_1+\tau x_i y_1)
L_{s-1}(\setminus x_1;\setminus y_1).
\end{equation}
Using the inductive step, we conclude that
\begin{equation}
\Res_{x_1=y_1^{-1}} L_s
=\Res_{x_1=y_1^{-1}} R_s.
\end{equation}
By symmetry, the residues at $x_1=y_j^{-1}$, $j=2,\ldots,s$, coincide as well.

Finally, it remains to check that $L_s$ and $R_s$ coincide at $x_1=0$.
For $R_s$, we have
\begin{equation}\label{R_0}
R_s\big|_{x_1=0} = 
\prod_{i=1}^s y_i 
\prod_{j=2}^{s} \prod_{k=1}^{s}
(x_j+y_k+\tau x_j y_k) \det_{1\leq j,k\leq s}
\left[ 
\begin{cases}
y_k^{-1} & j=1\\
\psi(x_j,y_k) & j\geq 2
\end{cases}
\right].
\end{equation}
For $L_s$ we use \eqref{ind-step}, where, when $x_1=0$, the sum over
$k$ reduces only to the term with $j=1$, because all the other terms
have the factor $x_1$, and we obtain
\begin{equation}
L_s\big|_{x_1=0} = \prod_{i=2}^n x_i \sum_{k=1}^s (-1)^{k+1}
\prod_{\substack{j=1\\ j\neq k}}^{s} y_j (1+y_j y_k+\tau y_k) 
R_{s-1}(\setminus x_1;\setminus y_k),
\end{equation}
whcih can be written in determinantal form
\begin{equation}\label{L_0}
L_s\big|_{x_1=0}
=\prod_{i=1}^s y_i 
\prod_{j=2}^{s} \prod_{k=1}^{s}
(x_j+y_k+\tau x_j y_k) 
\det_{1\leq j,k\leq s} 
\left[ 
\begin{cases}
v_k & j=1\\
\psi(x_j,y_k) & j\geq 2
\end{cases}
\right]
\end{equation}
with
\begin{equation}
v_k = y_k^{-1} \prod_{i=2}^{s} \frac{x_i}{x_i+y_k+\tau x_i y_k}
\prod_{\substack{j=1\\ j\neq k}}^{s}(1+y_j y_k+\tau y_k).
\end{equation}
As a function of $y_k$, the quantity $v_k$ vanishes as
$y_k\rightarrow \infty$ and has simple poles at $y_k=0$ and
$y_k=-x_j/(1+\tau x_j)$, $j=2,\ldots,s$. Therefore, the following pole
expansion is valid
\begin{equation}\label{polar}
v_k = y_k^{-1}-\sum_{j=2}^s 
\prod_{\substack{i=2\\ i\neq j}}^s \frac{x_i}{x_i-x_j}
\prod_{i=1}^s(1-x_j y_i)
\frac{1}{(1- x_j y_k)(x_j+y_k+\tau x_j y_k)}.
\end{equation} 
The last factor in \eqref{polar} can be recognized 
as the function $\psi(x_j,y_k)$. This implies equality 
of the determinants in \eqref{R_0} and \eqref{L_0},
and therefore
\begin{equation}
L_s\big|_{x_1=0}= R_s\big|_{x_1=0},
\end{equation}
that completes the proof of Proposition \ref{Cantini}.

\section{The Tracy-Widom relation}\label{appb}

In \cite{TW-08}, Tracy and Widom obtained the following
antisymmetrization relation (see \cite{TW-08}, equation (1.6)): 
\begin{multline}\label{TW}
\mathcal{A}_{z_1,\dots,z_s}
\left[\prod_{j-1}^s\frac{z_j^{s-j}}{1-\prod_{l=1}^jz_l}
\prod_{1\leq j<k\leq s}\left(qz_jz_k-z_k+p\right)\right]
\\
=p^{s(s-1)/2}\prod_{j=1}^s\frac{1}{1-z_j}\prod_{1\leq j<k\leq s}(z_j-z_k),
\end{multline}
where the parameters $p$ and $q$ obeying the condition $p+q=1$. 
In the context of the asymmetric simple
exclusion process $p$ and $q$ are 
the transition rates of the model.

Here we show that the antisymmetrization relation  \eqref{cantini}
reduces to \eqref{TW}, by setting
\begin{equation}\label{TWlim}
x_j=tz_j,\qquad y_j\to t^{-1},\qquad j=1,\dots,s,
\end{equation}
and
\begin{equation}\label{asep}
t=\sqrt{\frac{q}{p}},\qquad \tau =-\frac{1}{\sqrt{pq}}.
\end{equation}
The condition $p+q=1$, 
required for \eqref{TW} to hold, now reads 
\begin{equation}\label{condition}
t^2+\tau t+1=0.
\end{equation}
If we take $t$ as an independent parameter, then $\tau=-t-t^{-1}$.

Before turning to the calculation of both sides of \eqref{cantini} under 
the condition \eqref{condition} and in the limit \eqref{TWlim}, it is useful 
to note that these operations commute, as it follows from the 
formula \eqref{L-Poly}. It turns out that in 
evaluating the left-hand side it is convenient first to assume 
the condition \eqref{condition} and next perform the limit \eqref{TWlim}, but  
vice versa for the right-hand side.  

Considering the relation \eqref{cantini}, we divide it by
the Vandermonde product in $y$'s and perform the change of
the variables $x_j= tz_j$, $y_j= t^{-1}(1+\epsilon_j)$, $j=1,\dots,s$.
In the $\epsilon_j\to 0$ limit, $j=1,\dots,s$, the relation reads
\begin{multline}\label{cantinilim}
\lim_{\epsilon_1,\ldots,\epsilon_s\to0}\prod_{1\leq j<k\leq s}
\frac{1}{\epsilon_j-\epsilon_k}
\Asym_{z_1,\dots,z_s}\Asym_{\epsilon_1,\dots,\epsilon_s}
\Bigg[
\prod_{j=1}^s  \frac{[z_j(1+\epsilon_j)]^{s-j}}{1-\prod_{l=1}^j z_l(1+\epsilon_l)}
\\ \times
\prod_{1\leq j<k\leq s}(t^2 z_jz_k+\tau t z_k+1)
\left[1+\tau t^{-1}+t^{-2}+ t^{-2}\epsilon_j 
+\left(\tau  t^{-1}+t^{-2}\right)\epsilon_k+t^{-2}\epsilon_j\epsilon_k\right]
\Bigg]
\\ 
=W_s(t z_1,\dots,t z_s;t^{-1},\dots,t^{-1})\prod_{1\leq j<k\leq s}(z_j-z_k).
\end{multline}
Our aim is to evaluate both sides of this relation under the condition 
\eqref{condition}, that is, in the case where $\tau=-t-t^{-1}$ .

As for the left-hand side of
\eqref{cantinilim}, the evaluation of the limit is in general a cumbersome task, 
because all terms under the antisymmetrization symbol contribute to the leading order.  
If the condition \eqref{condition} is imposed,  
then only the last factor in the double product contributes.
Therefore, setting $\tau=-t-t^{-1}$, at leading order we have 
\begin{multline}\label{lhsepsilon}
\lim_{\epsilon_1,\ldots,\epsilon_s\to0}
\prod_{1\leq j<k\leq s}
\frac{1}{\epsilon_k-\epsilon_j}
\Asym_{z_1,\dots,z_s}\Asym_{\epsilon_1,\dots,\epsilon_s}
\Bigg[\prod_{j=1}^s  \frac{z_j^{s-j}}{1-\prod_{l=1}^j z_l}
\\ \times
\prod_{1\leq j<k\leq s}\left(t^2z_jz_k -(1+t^2) z_k+1\right)
\left(t^{-2}\epsilon_j-\epsilon_k\right)
\Bigg].
\end{multline}
Now, the antisymmetrization over $\epsilon_1,\dots,\epsilon_s$ can be
done explicitly.
\begin{lemma}\label{lemma3}
The following antisymmetrization relation is valid
\begin{equation}\label{smallid}
\Asym_{\epsilon_1,\dots,\epsilon_s}
\left[\prod_{1\leq j<k\leq s}\left(\epsilon_j-t^2\epsilon_k\right)\right]
=\prod_{j=1}^s\frac{1-t^{2j}}{1-t^2}
\prod_{1\leq j<k\leq s}(\epsilon_j-\epsilon_k).
\end{equation}
\end{lemma}
\begin{proof}
The result of the
antisymmetrization of a polynomial in $s$-variables of degree at most
$s-1$ is proportional to a Vandermonde product, and thus one just
needs to determine the overall constant. 
Denoting the left-hand side by $D_s=D_s(\epsilon_1,\dots,\epsilon_s)$, 
and considering its value as $\epsilon_s\to0$, we 
have 
\begin{align}
D_s\big|_{\eps_s=0}
&=\sum_{l=1}^s(-1)^{s-l}
\sum_{\substack{\sigma \\ \sigma_l=s}}(-1)^{[\sigma]}
\prod_{\substack{1\leq j<k\leq s\\ j,k \ne l}}
\left(\epsilon_{\sigma_j}-t^2\epsilon_{\sigma_k}\right)
\prod_{j=1}^{l-1}\epsilon_{\sigma_j}
\prod_{j=l+1}^s \left(-t^{2}\epsilon_{\sigma_j}\right)
\\ &
=D_{s-1}\prod_{j=1}^{s-1}\epsilon_j \sum_{l=1}^s t^{2(l-1)}.
\end{align}
Denoting $C_s$ the overall constant, $C_1=1$, we thus get 
\begin{equation}
C_s=C_{s-1}\sum_{l=1}^s t^{2(l-1)}=C_{s-1}\frac{1-t^{2s}}{1-t^2}
=\prod_{j=1}^{s}\frac{1-t^{2j}}{1-t^2},
\end{equation}
and \eqref{smallid} follows.
\end{proof}
As a result, we get the following expression  
\begin{equation}\label{lhslimit}
\frac{1}{t^{s(s-1)}}\prod_{j=1}^s\frac{1-t^{2j}}{1-t^2} 
\Asym_{z_1,\dots,z_s}
\Bigg[\prod_{j=1}^s\frac{z_j^{s-j}}{1-\prod_{l=1}^j z_l} 
\prod_{1\leq j<k\leq s}(t^2 z_j z_k -(1+t^2) z_k+1) \Bigg]
\end{equation}
for the left-hand side of \eqref{cantinilim} under 
the condition \eqref{condition}.

As for the right-hand side of \eqref{cantinilim},
it turns out that the determinant in \eqref{defW} can be evaluated 
explicitly, provided that \eqref{TWlim} and \eqref{condition} are fulfilled.
Consider first the function $W_s(x_1,\ldots,x_s,y_1,\ldots,y_s)$
in the limit where all $y_1,\ldots,y_s$ tend to the same value $y$, while 
$x_1,\ldots,x_s$ remain different from each other. A standard calculation gives
\begin{equation}\label{Wxy}
W_s(x_1,\dots,x_s;y,\dots,y)
=\frac{\prod_{j=1}^{s}
(x_j+y+\tau x_j y)^s}{\prod_{1\leq j<k \leq s}(x_k-x_j)}
\det_{1\leq j,k \leq s}  \left[\frac{1}{(k-1)!}\partial_y^{k-1}\psi(x_j,y)\right].
\end{equation}
Using for the function $\psi(x,y)$ the identity 
\begin{equation}
\frac{1}{(1-xy)(x+y+\tau x y)}
=
\frac{1}{1+\tau x +x^2}\left\{\frac{1}{y+x(1+\tau x)^{-1}}
-\frac{1}{y-x^{-1}}\right\},
\end{equation}
we get
\begin{equation}
\frac{1}{(k-1)!}\partial_y^{k-1}\psi(x,y)=
\frac{(-1)^{k-1}}{1+\tau x +x^2} \left\{\frac{1}{\left(y+x(1+\tau x)^{-1}\right)^{k}}-
\frac{1}{\left(y-x^{-1}\right)^k}\right\}.
\end{equation}
Now, setting in the last expression $x=tz$, $y=t^{-1}$, and $\tau=-t-t^{-1}$,  
we obtain 
\begin{equation}
\begin{split}
\frac{1}{(k-1)!}\partial_y^{k-1}\psi(tz,y)\Bigg|_{
\begin{subarray}{l}
y=t^{-1}\\
\tau=-t-t^{-1}
\end{subarray}}
&=
\frac{(-1)}{(1-z)(1-t^2 z)} 
\left\{\frac{t^k[1-(1+t^2)z]^k}{(1-z)^k}-\frac{(tz)^k}{(z-1)^{k}}\right\}
\\ &
=\frac{t^k\{z^k-[(1+t^2)z-1]^k\}}{(1-z)^{k+1}(1-t^2 z)}. 
\end{split}
\end{equation}
Therefore, applying \eqref{TWlim} and \eqref{condition} to \eqref{Wxy} yields  
\begin{multline}\label{Ds}
W_s(t z_1,\dots,t z_s;t^{-1},\dots,t^{-1})\Big|_{\tau=-t-t^{-1}}
=\frac{1}{t^{s(s-1)}}\prod_{1\leq j<k \leq s}\frac{1}{z_k-z_j}\prod_{j=1}^{s}
\frac{1}{1-z_j}
\\
\times  \det_{1\leq j,k \leq s}  
\left[(1-z_j)^{s-k}\frac{z_j^k-[(1+t^2)z_j-1]^k}{1-t^2 z_j}\right].
\end{multline}
The determinant in \eqref{Ds} is proportional to the Vandermonde product, 
because the entries of the matrix have the form $P_{k}(z_j)$, where 
$P_{k}(z)$, $k=1,\ldots,s$, are all polynomials of degree $s-1$. 
The overall constant can be fixed by 
considering the determinant, for example, at $z_s=1$, along the lines of 
Lemma~\ref{lemma3}. In this way, we obtain 
\begin{equation}\label{DsVan}
\det_{1\leq j,k \leq s}  
\left[(1-z_j)^{s-k}\frac{z_j^k-[(1+t^2)z_j-1]^k}{1-t^2 z_j}\right]
=\prod_{j=1}^{s} \frac{1-t^{2j}}{1-t^2}
\prod_{1\leq j<k \leq s}(z_k-z_j),
\end{equation}
and hence
\begin{equation}
W_s(t z_1,\dots,t z_s;t^{-1},\dots,t^{-1})\big|_{\tau=-t-t^{-1}}
=\frac{1}{t^{s(s-1)}}
\prod_{j=1}^s\frac{1-t^{2j}}{1-t^2} \prod_{j=1}^s\frac{1}{1-z_j}.
\end{equation}
As a result, we get the following expression 
\begin{equation}\label{rhslimit}
\frac{1}{t^{s(s-1)}}
\prod_{j=1}^s\frac{1-t^{2j}}{1-t^2} \prod_{j=1}^s\frac{1}{1-z_j}
\prod_{1\leq j<k\leq s}(z_j-z_k)
\end{equation}
for the right-hand side of \eqref{cantinilim} under the condition 
\eqref{condition}.

Finally, equating \eqref{lhslimit} and  
\eqref{rhslimit} and setting $t=\sqrt{q/p}$, with $p+q=1$, we 
arrive at the relation \eqref{TW}. 

\bibliography{cantide_bib}
\end{document}